# Automatic Detection of Reflective Thinking in Mathematical Problem Solving based on Unconstrained Bodily Exploration

Temitayo A. Olugbade, Joseph Newbold, Rose Johnson, Erica Volta, Paolo Alborno, Radoslaw Niewiadomski, Max Dillon, Gualtiero Volpe, Nadia Bianchi-Berthouze

**Abstract**—For technology (like serious games) that aims to deliver interactive learning, it is important to address relevant mental experiences such as reflective thinking during problem solving. To facilitate research in this direction, we present the weDraw-1 Movement Dataset of body movement sensor data and reflective thinking labels for 26 children solving mathematical problems in unconstrained settings where the body (full or parts) was required to explore these problems. Further, we provide qualitative analysis of behaviours that observers used in identifying reflective thinking moments in these sessions. The body movement cues from our compilation informed features that lead to average F1 score of 0.73 for binary classification of problem-solving episodes by reflective thinking based on Long Short-Term Memory neural networks. We further obtained 0.79 average F1 score for end-to-end classification, i.e. based on raw sensor data. Finally, the algorithms resulted in 0.64 average F1 score for subsegments of these episodes as short as 4 seconds. Overall, our results show the possibility of detecting reflective thinking moments from body movement behaviours of a child exploring mathematical concepts bodily, such as within serious game play.

—————————— ◆ ——————————

## 1 INTRODUCTION

THERE is a consensus in education literature that reflective thinking is integral to learning [1][2][3][4]. Findings in [5] suggest that this cognitive strategy may in fact be necessary to solution of mathematical problems. Indeed, the authors demonstrated the possibility of guiding a learner through the use of reflective thinking in solving mathematical problems. Like a competent human teacher would [6], digital learning technology should be capable of providing personalised support to foster application of this strategy [7]. To deliver such personalisation, it is essential for the technology to be able to detect when the learner is (not) thinking reflectively in solving the given problem. To address this, the current paper reports our investigation of the possibility of automatic detection of reflective thinking in the context of mathematical problem solving, toward technological personalisation aimed at promoting use of reflective thinking strategy.

We frame the problem in the context of the weDraw serious games [8] in which children explore mathematical concepts (e.g. angle arithmetic) based on bodily interaction. This design is grounded in findings in child education literature that children embody knowledge of and reflection on mathematical concepts [9][10]. With reflective thinking detection capability, each weDraw game could, for instance, instantaneously adapt the time it provides for a child to solve a given problem, with the aim of allowing them time to think reflectively in solving the problem. Each game could additionally provide prompts that help the child relate the problem to previous knowledge, tailoring when, how, and what (a) prompt is delivered to its detection of reflective thinking in the child. Consider the example of an angle arithmetic problem where the child is to find the sum of two angles diagrammatically (rather than using their numeric values). The child should be given adequate time to explore the problem without interruption [11]. If after such exploration, the child struggles with finding the solution, the game could ask whether they expect the solution to be larger or smaller than the bigger of the two angles. If the child arrives at the correct answer to this question and the game detects that this was done thinking reflectively, the game could continue to provide further cues, e.g. getting the child to perhaps remember how much larger a familiar object (such as hands) will get when it is joined with others, until the child arrives at the correct solution.

Given the settings of the weDraw games, i.e. bodily interaction of children in mathematical problem solving, we

---

- *T.A. Olugbade is with the University College London, London, WC1E 6BT. E-mail: temitayo.olugbade.13@ ucl.ac.uk.*
- *J. Newbold is with the University College London, London, WC1E 6BT. E-mail: joseph.newbold.14@ ucl.ac.uk.*
- *R. Johnson is with the University College London, London, WC1E 6BT. E-mail: rose.johnson@ ucl.ac.uk.*
- *E. Volta is with the University of Genoa, Genoa, 16126. E-mail: er-ica.volta@edu.unige.it.*
- *P. Alborno is with the University of Genoa, Genoa, 16126. E-mail: paoloalborno@gmail.com.*
- *R. Niewiadomski is with the Italian Institute of Technology, Genoa, 16163. E-mail: radoslaw.niewiadomski@iit.it.*
- *M. Dillon is with PwC, 7 More London Riverside, SE1 2RT. E-mail: max.dillon@pwc.com.*
- *G. Volpe is with the University of Genoa, Genoa, 16126. E-mail: gualtiero.volpe@unige.it.*
- *N. Bianchi-Berthouze is with the University College London, London, WC1E 6BT. E-mail: nadia.berthouze@ ucl.ac.uk.*

investigate automatic detection of reflective thinking based on bodily cues during mathematical problem solving with children. The long-term aim is to create adapting, movement-based mathematical games that deliberately support children in using reflective thinking strategies to explore new problems. Our main contributions in the current paper are as follows:

- we build a novel annotated dataset (named weDraw-1 Movement Dataset) on body movement of children exploring mathematical concepts in unconstrained settings. One portion of the dataset (details in Section 4) was collected in a school and so has additional complexity that more closely matches the intricacies of normal classroom lessons where there are space, time, and setup constraints.

- we provide the first known in-depth analysis of the reflective thinking behaviours of children during mathematical problem solving, based on the weDraw-1 Movement Dataset.

- we contribute understanding of how reflective thinking periods can be modelled from body movement data, with a focus on the use of Long Short-Term Memory neural networks (LSTMNN) and based on the weDraw-1 Movement Dataset.

The paper is organized as follows: in Section 2, we provide a background on reflective thinking, and discuss the state of the art in the detection of learning-related mental states in general in Section 3. The acquisition of our weDraw-1 Movement Dataset (Study 1) is reported in Section 4 while the analysis of reflective thinking behaviours (Study 2) is discussed in Section 5. In Section 6, we describe our investigations of binary classification of problem-solving periods by reflective thinking (Study 3). Our findings are altogether discussed in Section 7, and a conclusion is provided in Section 8.

## 2 BACKGROUND: A DEFINITION OF REFLECTIVE THINKING

In this section, we discuss the definition of reflective thinking that we used for the annotation of the weDraw-1 Movement Dataset based on observer ratings. This annotation was necessary for obtaining the ground truth for our automatic detection investigations.

Although reflective thinking is widely mentioned in education literature (as 'reflection' or 'reflectivity'), there is no straightforward definition useful for characterising it in the context of problem solving. This may be because the term is most commonly used within the context of experiential learning to refer to post-activity reflection [1][2][3]. The most appropriate definition found for problem solving contexts is the classic definition of Dewey [4]:

"… turning a subject over in the mind and giving it serious and consecutive consideration [p. 3]. In between [pre- and post-reflective periods, i.e. during the reflective thinking process] … are (1) suggestions, in which the mind leaps forward to a possible solution; … and (5) testing the hypothesis by overt or imaginative action [p. 107]." (pp. 3, 107)

Dewey [4] further differentiates reflective thinking from other forms of 'thinking', arguing that reflective thinking is particularly preceded by doubt, mental difficulty, or perplexity, and characterised by searching and inquiring aimed at resolving these. He also stresses a distinction between reflective thinking and merely reaching a conclusion (or producing an answer) without critically testing the ideas or solutions that emerge in the mind.

Still, Dewey's treatise on reflective thinking [4] does not provide clear directives on how it can be recognised, whereas a clear definition is important for observer annotation [12]. The limitation of Dewey's definition is inherent to the process of reflective thinking itself, which is essentially internal [4]. On one hand, the provision of an answer or solution that is correct does not in itself help an observer judge that reflective thinking has taken place in solving the problem [4]. This is because the learner could provide the correct answer due to familiarity with the problem which makes a solution handy. The correct answer could also be provided by chance. On the other hand, an answer or solution that is incorrect, or even no solution arrived at, is not on its own helpful to the observer. Although Dewey [4] implies that reflective thinking always leads to a settled and harmonious state (and correct solution), we rather surmise that a fault along the pipeline may lead to a confused or a wrongly confident state. Thus, the observer has to rely on other cues to judge if the learner arrived at the solution (right, wrong, or none) through reflective thinking. To aid such judgement of reflective thinking from behaviour during problem solving, we unfurl Dewey's definitions, highlighting two factors critical for supposing reflective thinking and resulting in a definition of observation of reflective thinking as:

> _observation_ that the learner _takes time to consider a problem or its solutions_ [_whether in search of an appropriate solution due to: unfamiliarity with the problem, critiquing of alternatives despite having a solution ready (or critiquing of the solution itself), or some other analytical approach that contributes to solving the problem_].

## 3 RELATED WORKS: AUTOMATIC DETECTION OF LEARNING-RELATED STATES

The only study to have previously investigated automatic detection of thinking in learning settings is [13] who considered thinking and five other self-reported states in adults, within a seated PC-based learning context. The authors found thinking to be the most frequently occurring of these states, reinforcing the importance of addressing reflective thinking within digital learning technology. Using a dynamic Bayesian model based on upper body gestures (chin rest, head scratch, eye rub, lip touch, locked fingers, and yawn), they achieved accuracy of 0.97. While the model is limited by dependence on initial recognition of these gestures (performed manually in their paper), the finding suggests that automatic detection of reflective thinking is feasible from bodily expressions within a constrained setting. Their analysis further highlighted chin rest and lip touch as the gestures most indicative of think-ing. A related study is the work of Bosch et al. [14] in which face video data was captured in learning tasks in a group setting. Although each of the 20 students in a group session was constrained to individual desks during the PC-based learning tasks, the students were free to interact with other students, e.g. in discussing the given tasks. Exploring several techniques (clustering, Naives Bayes, and Bayes Net) on extracted features, the authors obtained 0.69 accuracy on average for binary classification of bored, confused, delighted, engaged, and frustrated. These studies are representative of the work in the area of learning-related affect detection (see, e.g., [7][15] for other review) which has until now focused on settings where the learner is constrained, e.g. to interaction using a mouse/keyboard. Our study aims to extend these findings by focusing on unconstrained settings where interaction relies on bodily gestures and the learner is free to move around and interact with objects in the space. The child is additionally not restrained from interaction with the instructor or parent or siblings (e.g. chatting with a sibling about how the given problem could be solved) who were sometimes in the room during the problem-solving sessions. This context presents a greater challenge than constrained PC-based settings. Further, we propose a model that more directly uses sensor data (and so accesses richer information) and with no need of manual annotation of individual expressions.

Another relevant work is the study in [16] on automatic detection of engagement (both behavioural and affective) in kindergarten age children during seated learning interaction with a PC, based on neural network modelling. The ground truth was obtained using both expert and naïve observers who reached good level of agreement (0.62 Cohen's kappa for both) suggesting that naïve observers can provide reliable ratings of learning-related states in child subjects. In Yun et al.'s network architecture [16], the first layers, based on a deep pretrained CNN, form an image-level processor. The outputs from these layers are max-pooled and concatenated for a video sequence and fed into temporal processing layers which comprise a parallel configuration of convolutional cum max pooling, max pooling, average pooling, and variance pooling layers. The outputs from this portion are concatenated and finally input into a series of convolutional, fully connected, rectified linear unit, dropout, and fully connected layers. Using this model for binary discrimination of engagement, with 3907 sequences, Yun et al. [16] obtained accuracy of 0.81 (Matthews correlation coefficient of 0.52). Although the data type in their work (videos) differs from our own focus (motion capture data), their findings point to value in exploring a multilayer architecture where learning is done progressively across dimension scales (e.g. from single image level to video level). This approach has been around for a while. For example, it has been explored for action recognition [17] and it is not far removed from the longstanding techniques of multimodal data fusion in the area of affective computing (see [18] for a review). We look beyond convolutional layers typical for computer vision problems. In particular, we explored LSTM layers, which are designed to process sequences, such as the timeseries in motion capture data [19][20]. Findings in body movement analysis in the context of activity recognition suggest that the LSTM layer can also learn by integration across temporal and non-temporal dimensions [21][22]. However, previous studies consider well-defined gestures (e.g. door opening) or alteration from well-defined trajectories (e.g. deviation from expected manoeuvre during execution of a physical exercise) [21] rather than higher level and subjective interpretation of bodily cues, e.g. reflective thinking.

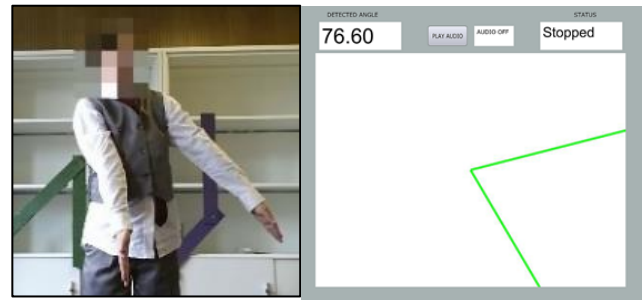

Fig. 1. (a) Left - A static representation of 45° using the arms, in the Forming Angles task; (b) Right - Visual feedback for a static representation of 75° using the arms, in the same task.

We discuss our investigations in the next sections. First, we describe the weDraw-1 Movement Dataset collected to facilitate our investigations and discuss our analysis of behavioural cues of reflective thinking that emerged in this dataset. We then present our modelling of reflection thinking periods.

## 4 STUDY 1: WEDRAW-1 MOVEMENT DATASET COLLECTION

The weDraw-1 Movement Dataset was collected in two main settings. In Setting A, we used a room within the university and there was one of the child's parents (and sometimes one or two siblings) present in addition to the child and two researchers, one of whom interacted with the child and acted as the instructor. Setting B was a (smaller) room within the primary school that the child was attending at the time of the study, with only the two researchers and the child present.

All of the data were collected in the UK and all of the children were studying in the UK at that time, in school years between Year 2 and Year 7 with an average of 4.38 (standard deviation of 1.47). The data was collected from a total of 26 children (14 children in Setting A and 12 in Setting B) between 6 and 11 years old with mean age of 8.69 years and standard deviation of 1.19.

The dataset comprises 120 sequences (64 in Setting A and 56 in Setting B) of video with corresponding three-dimensional full-body positional data, each of a child performing a single task or multiple related tasks. The sequences are from 24 children whose parental consent permitted us to get their videos annotated. The longest sequence is 537.23 seconds in length (median = 117.94, interquartile range = 154.18). Both the video and positional data were captured using the Microsoft Kinect v2 sensor.

## 4.1 Mathematical Problem-Solving Tasks

To support the core idea of the weDraw project, which is centred on bodily exploration of mathematical concepts, the tasks in which the data was collected were set up to encourage considerable exploration using the body. The tasks were further designed to create real learning experiences and several pilot tests were carried out to iteratively improve the interaction and exploration experience of the child participants. The problems in the tasks were informed by pedagogical studies partially published in [23], with a focus on angles, symmetry, and shape reflection. There were five main types of task (these tasks were performed with repetitions, usually completed in the given order):

- **Forming Angles**: In this task type, the child explored static representation of given angles using their arms (e.g. Fig. 1a) and received automatic visual feedback (on a screen) based on an early prototype of one of the weDraw games. The visual feedback (see Fig. 1b) consists of a numerical value and a diagrammatic representation of the angle formed by the child's arms, in real-time.

- **Bodily Angles Sums and Differences**: Here, the child was given a pair of angles, each represented by a three-dimensional object (named 'Angle-Arms') depicting the rays of an angle, e.g. 135° and 45° in Fig. 2a. The Angle-Arms were attached to a wall and the task for the child was to represent with his/her own arms the angle resulting from the sum (or difference) of the given angles. One strategy to solve the problem would require the child to first place his/her arms against the rays of the first Angle-Arms to represent its angle. The child would then need to go to the second Angle-Arms and align one of his/her arms against one of the Angle-Arms's rays, keeping the angle representation of the first Angle-Arms. After this, the child would have to sweep his/her aligned arm (keeping the other still) towards the second ray of this second Angle-Arms. To reduce the complexity of the problem, the Angle-Arms were arranged such that a ray of the first had the same orientation as a ray of the second (see example in Fig. 2a).

  The problem was adapted for the youngest children (i.e. children aged 6 years old). These children were given Angle-Arms pairs or trios (two either of the same magnitude or of only 10° difference in magnitude) and asked to find the largest (or smallest) angle. The Angle-Arms were arranged such that the rays all had different orientations, e.g. Fig. 2b. The child was encouraged to use his/her body to explore the angles in finding the solution.

- **Rotating in Angles:** Each child was asked to represent the sum/difference obtained in Bodily Angles Sums and Differences tasks as full-body rotation. Lines on the floor dividing an imaginary circle around the child's feet into octants (see Fig. 3a) were used to provide visual guide for this task.

- **Finding Symmetry**: Here, the child was seated in front of a table and asked to choose from a set of large cardboard shape cut-outs (diameter between 35 and 45.5cm) and show the lines of symmetry of the shape, if any. The younger children were usually first asked to explore the basic shape properties: the number of edges and vertices, and the name.

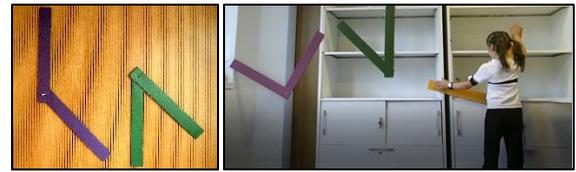

Fig. 2. (a) Left - 135° and 45° Angle-Arms attached to the wall, for the Bodily Angles Sums task; (b) Right - A 6-year old child gauging the size of a 90° Angle-Arms using her arms, in an adapted version of the same tasks, with two additional Angle-Arms: another 90° and a 45°.

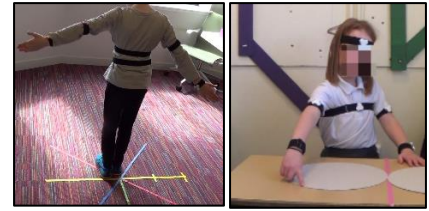

Fig. 3. (a) Left - A snapshot of a child performing a Rotating in Angles task, the coloured tapes on the floor around the child's feet divided an imaginary circle into octants; (b) Right - A snapshot of a child solving a Making Shape Reflections problem.

- **Making Shape Reflections**: In this case, the child was given a duplicate of the chosen shape and asked to arrange the two cutouts such that one was a reflection of the other, with as many reflection configurations as possible. A line taped across the table (see Fig. 3b) was used to simulate a mirror.

## 4.2 Data Annotation

As the weDraw serious games are envisioned as teaching systems that support traditional learning settings with human teachers, it is practical to model the proposed automatic detection system after human assessment. Thus, similar to [16], we obtained human observer assessment for ground truth for automatic detection of reflective thinking. Although findings [17] suggest that untrained observers can provide reliable ratings of learning-related states in children, pilot studies that we carried out with 8 observers (4 female) with teaching/tutoring experience and psychology students/experts suggest that this may not translate to complex learning settings like ours. Thus, two of our researchers (R1 and R2) with experience working with children and also present during the data collection (and so familiar with the tasks given to the children) independently labelled the data. The raters continuously rated all 120 videos recorded (without audio, to force them to rely on visual cues, similar to [24]), using the Elan annotation software [25][26]. The raters specifically marked periods of reflective thinking within these videos. The raters also marked periods of low confidence, which is another cognitive state of interest in learning settings [7]; however, we focus on the reflective thinking labelling in this paper.

TABLE 1
LIST OF CODES THAT EMERGED FROM THE ANALYSIS OF REFLECTIVE THINKING BEHAVIOURAL CUES

| Code | Examples | Higher Level Code |
|---|---|---|
| Speak to self | | Verbal |
| Smile | | Facial, Mouth |
| Speech delay | "opens mouth as if to speak, but not speaking" (R1); "seeming to utter 'uhm'" (R1) | Facial, Mouth |
| Doing something to the mouth with the mouth | "pushes lips upwards and release just before response" (R1) | Facial, Mouth |
| Other mouth expression | | Facial, Mouth |
| Doing something to the eye with the eye | "beginnings of a frown" (R1); "squints eyes" (R1) | Facial |
| Other facial expression | "hard look" (R2) | Facial |
| Finger(s) touching head region | "hands clapped over nose" (R1); "scratches head" (R1); "finger to mouth" (R2); "rubs head" (R2) | Body (hand) |
| Pointing | | Body (hand) |
| Head tilt | | Gaze |
| Looking into space/ground/ceiling | | Gaze |
| Looking at relevant object (e.g. shape cut-outs), while in non-action | | Gaze |
| Other gaze (change) | | Gaze |
| Forward lean | | Body (trunk) |
| Back lean | | Body (trunk) |
| Pause at the start | | Body (whole); Time |
| Pause at the end | | Body (whole); Time |
| Other pause | | Body (whole); Time |
| Problem solving duration | | Time |
| Slow movement | | Time; Body (multiple) |
| Tentative/cautious movement | | Body (multiple) |
| Fidget | "worries mouth" (R1); "swings … leg, … stops, … starts again" (R1) | Body (multiple) |
| Other gesture/posture | | Body (multiple) |
| Reminds self of problem/question | | |
| Exploration by own movements | | Exploration |
| Exploration by moving object | | Exploration |
| Systematic solving | | |
| Solution implementation or response | | |
| Gesturing while speaking | | |
| Other | | |

They reviewed their annotations several times, based on the definition discussed in Section 2 and after consulting with one another (such as done in [27]).

To understand how much the raters agreed on the occurrence of reflective thinking, we computed a two-way mixed model, absolute agreement, average measures intraclass correlation (ICC) on the labels for the Setting A data (249,126 frames). Similar to the approach of Griffin et al. [28], we accounted for expected misalignment in the onset and offset of periods which the two raters agree are reflective thinking periods. This was done in our work by adjusting the rater labels such that overlaps of positive labels (between R1 and R2) were synchronised, with the earliest onset and the latest offset assumed for both raters. We found ICC = 0.63, which shows good level agreement [29].

During the annotation, after marking a time period as a moment of reflective thinking, the raters further noted the cues that they used in recognising reflective thinking at that specific moment. The primary aim of this was to inform the extraction of hand-crafted features for automatic detection of reflective thinking. We analysed the reported cues; the analysis of the cues reported for the Setting A data is discussed in Section 5.

## 5   STUDY 2: REFLECTIVE THINKING BEHAVIOUR ANALYSIS

### 5.1 Analysis Method

There were 531 cue reports for the Setting A data; each report specified a sequence or concurrence of behaviours, e.g. "… she pauses and then makes a thoughtful expression with her lips then looks up and away and then draws the answer with her finger …" (R2).

Based on thematic analysis methods [30], these cues were coded, and the codes were refined until all codes were clearly defined and no new themes emerged.

### 5.2 Findings

Table 1 shows the list of codes that emerged, with examples from associated cue report extracts.

Although the majority of these codes highlighted bodily behaviours, facial expressions, gaze and verbal behaviours were also noted. It was interesting that the raters (both of them) used a verbal cue in their judgments even though they were not provided with aural data. Indeed, the cue, speaking to oneself (i.e. private speech [31]), is known to be an observable behaviour and employed by children in a self-regulatory role while solving challenging problems [31].

Some of the facial cues used by the raters seemed difficult to specify in vernacular, e.g. "hard look" (R2), "thinking face" (R1). However, the majority of the facial codes are related to cues involving the mouth. Perhaps, these mouth behaviours serve to communicate that expected speech is delayed because the speech content is being generated or processed (in reflective thinking). Other facial expressions involved the eye. We theorise that this class of expressions and the gaze behaviour of looking away may function as a means of avoiding visual distractors and focusing attention on internal processes (e.g. recollection [4]), to solve the problem at hand. The link between thinking and looking away has been previously noted [32]. Head tilting was another gaze behaviour highlighted by the raters. This expression may, together with the forward/backward lean and exploration by moving the object to be manipulated also noted, have permitted the child a different visual perspective of the problem. Exploration may have additionally served in physically evaluating generated ideas.

It is not surprising that tempo (pauses, speed, duration) were used by the raters to recognise reflective thinking since this is the one cue specified in the definition of the construct (see Section 2). What is interesting to note from the cues reported is that reflective thinking seemed to occur both at the start and end of problem solving. Another intriguing note is that certain forms of fidgeting seem to accompany reflective thinking. It is not clear what the function of these behaviours is, but they may be used to fill the pause characteristic of reflective thinking or perhaps to break the stillness of that pause so as to stimulate ideas.

Other bodily cues highlighted by the raters involved self-touching of regions on the head. As discussed in Section 3, similar behaviours (chin rest, lip touch) have been previously associated with thinking [13].

Providing a solution was also a cue used; although the raters had agreed not to rely on this, it was decided that for some of the tasks, the motion of completing the task and arriving at the final solution could count toward evidence of reflective thinking. The manner in which the solution is described to the instructor, particularly elaborate use of gestures, was also used to infer reflective thinking.

## 6   STUDY 3: BINARY CLASSIFICATION OF PROBLEM-SOLVING PERIODS BY REFLECTIVE THINKING

In this section, we report our investigation of the modelling of reflective thinking periods using LSTMNNs.

As discussed in Section 3, we focus on LSTMNNs because of their inherent capability to learn temporal patterns. We specifically used bi-directional LSTM layers [33] so as to learn both forward and backward chains of the body movement events in the (not) reflective thinking periods. The rationale for this is based on our experience (during the analysis discussed in Sections 4.2 and 5) that an observer's interpretation of past movement events may be informed by current movement events (i.e. retrospection).

In the rest of this section, we describe the LSTMNN architectures explored, the input data used, and the findings of our investigation.

### 6.1 Dimension-Distributed and Vanilla LSTMNN Architectures

We explored two LSTMNN architectures: Dimension-Distributed and Vanilla bi-LSTMNNs (DD-LSTMNN and V-LSTMNN respectively). The two architectures are illustrated in Fig. 4. The DD-LSTMNN, where a shared bi-LSTM function is performed separately for each dimension of the input before being integrated using a fully connected layer (with rectified linear activation), is a simple architecture that draws on the multiscale approach in networks such as Yun et al.'s [16] (see discussion in the Section 3). In the V-LSTMNN, on the other hand, the bi-LSTM function is computed collectively for all dimensions.

For both architectures, we systematically experimented with different network depths, each with leave-one-subject-out cross-validation. We found one and two hidden layers to be the optimum for the DD- and V-LSTMNNs respectively based on our dataset.

### 6.2 Input Data

The full-body positions and reflective thinking labels (from rater R1) of the Setting A and Setting B data of the weDraw-1 Movement Dataset (see Section 4) were used to investigate the automatic detection of reflective thinking. Two of the 64 sequences in the Setting A data and one of the 56 in the Setting B data were not included in our investigation due to unavailability of positional data as a result of technical malfunction during data capture.

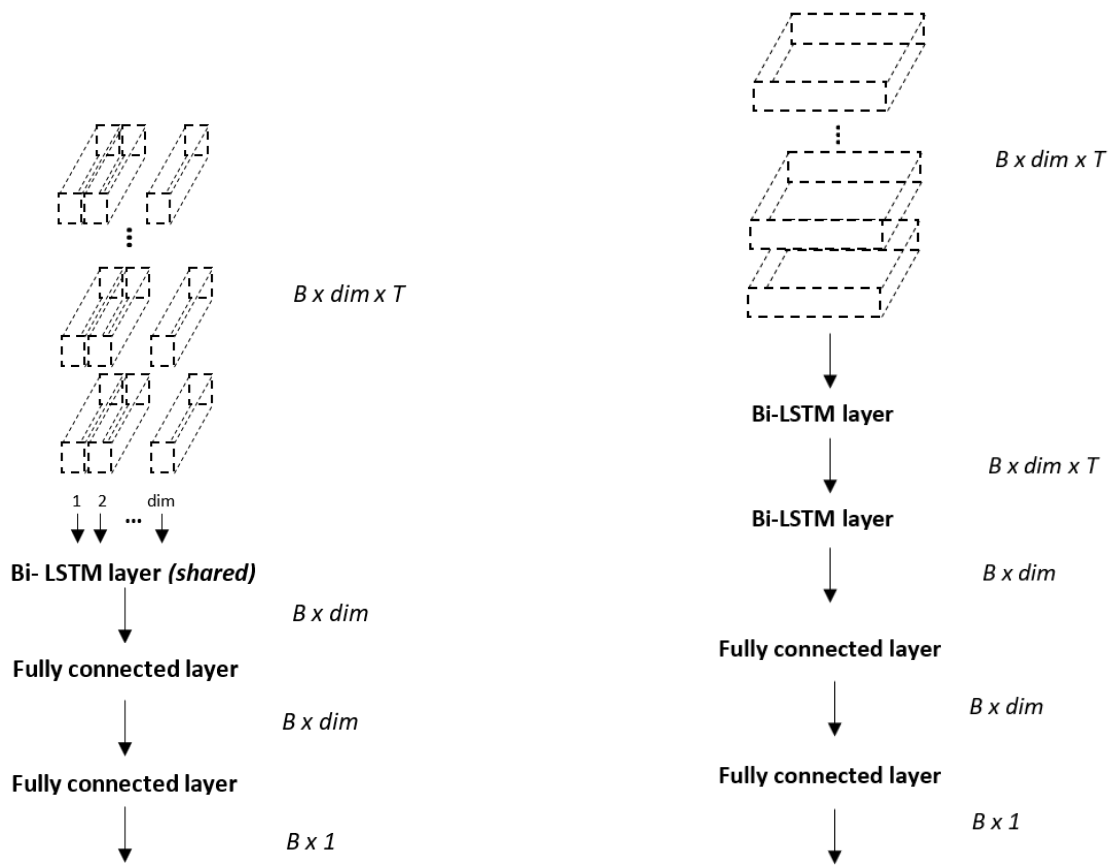

Fig. 4. (a) Left - Dimension-Distributed bi-LSTMNN; (b) Right - Vanilla bi-LSTMNN.

TABLE 2
FREQUENCY DISTRIBUTION OF REFLECTING THINKING (RT) AND NOT REFLECTIVE THINKING (NRT) ACROSS TASKS

| Frequency | Setting A | | | | Setting B | | | |
| --- | --- | --- | --- | --- | --- | --- | --- | --- |
| | Finding Angles | Bodily Angles Sums & Rotating in Angles | Bodily Angles Differences & Rotating in Angles | Finding Symmetry & Making Shape Reflections | Finding Angles | Bodily Angles Sums & Rotating in Angles | Bodily Angles Differences & Rotating in Angles | Finding Symmetry & Making Shape Reflections |
| RT | 15 | 81 | 17 | 83 | 29 | 60 | 4 | 72 |
| NRT | 15 | 81 | 16 | 83 | 29 | 60 | 4 | 72 |
| Duration (frames) | 19560 | 55050 | 8179 | 73475 | 36659 | 61284 | 1495 | 75526 |

In the rest of this section, we describe the segmentation of these data, augmentation of the data, two input forms derived from the data, and the approach used to deal with variation in segment lengths.

### 6.2.1 Data Segmentation

To prepare the data for modelling, we first split the full-body positional data by task per child and then segmented each task data $S_b$ ($b$ = 1, 2, …, n sequences; n=62 and 55 for Setting A and B data respectively) by periods of contiguous reflective thinking positive (or negative) frames in the corresponding label data. Periods at the end of each sequence and negative for reflective thinking (NRT) were excluded so as to balance with the number of periods positive for reflective

TABLE 3
29 FEATURES EXTRACTED AS INPUT DATA (POS) FOR AUTOMATIC DETECTION

(NUMBERING IN SUPERSCRIPT; $j_{i,t}^{b,m}$, IS WRITTEN AS $j_{i_t}$ FOR CONVENIENCE)

| Feature | Formula |
|---|---|
| Head twist/lateral-bend[1] and flexion[2] | $\tan^{-1}\left(\dfrac{\left\|(j_{head_t} - j_{neck_t}) \times (j_{i1_t} - j_{i2_t})\right\|}{(j_{head_t} - j_{neck_t}) \cdot (j_{i1_t} - j_{i2_t})}\right),$ $for\ (j_{i1_t}, j_{i2_t}) = (j_{lshoulder_t}, j_{rshoulder_t})\ and\ (j_{midspine_t}, j_{spinebase_t})\ respectively$ |
| Trunk flexion (left[3] and right[4] hand sides) | $\tan^{-1}\left(\dfrac{\left\|(j_{topspine_t} - j_{midspine_t}) \times (j_{i1_t} - j_{i2_t})\right\|}{(j_{topspine_t} - j_{midspine_t}) \cdot (j_{i1_t} - j_{i2_t})}\right),$ $for\ (j_{i1_t}, j_{i2_t}) = (j_{spinebase_t}, j_{lhip_t})\ and\ (j_{spinebase_t}, j_{rhip_t})\ respectively$ |
| Positional energy of the head[5], left and right hand[6,7] and knee[8,9] | $\dfrac{(j_{i_t} - j_{i_{t-1}})^2}{2},\ for\ j_{i_t} = j_{head_t}, j_{lhand_t}, j_{rhand_t}, j_{lknee_t}, j_{rknee_t}\ respectively$ |
| Angular energy of neck[10], left and right shoulder[11,12], elbow[13,14], hip[15,16], and knee[17,18] | $\dfrac{(a_{i_t} - a_{i_{t-1}})^2}{2},$ $where\ a_{i_t} = \tan^{-1}\left(\dfrac{\left\|(j_{i1_t} - j_{i0_t}) \times (j_{i2_t} - j_{i0_t})\right\|}{(j_{i1_t} - j_{i0_t}) \cdot (j_{i2_t} - j_{i0_t})}\right),$ $for\ (j_{i1_t}, j_{i0_t}, j_{i2_t})$ $= (j_{head_t}, j_{neck_t}, j_{topspine_t}), (j_{topspine_t}, j_{lshoulder_t}, j_{lelbow_t}), (j_{topspine_t}, j_{rshoulder_t}, j_{relbow_t}),$ $(j_{lshoulder_t}, j_{lelbow_t}, j_{lhand_t}), (j_{rshoulder_t}, j_{relbow_t}, j_{rhand_t}),$ $(j_{spinebase_t}, j_{lhip_t}, j_{lknee_t}), (j_{spinebase_t}, j_{rhip_t}, j_{rknee_t}),$ $(j_{lhip_t}, j_{lknee_t}, j_{lankle_t}),\ and\ (j_{rhip_t}, j_{rknee_t}, j_{rankle_t})\ respectively$ |
| Hand-to-head distance[19] | $\min(\{\|j_{head_t} - j_{lhand_t}\|, \|j_{head_t} - j_{rhand_t}\|\})$ |
| Range of movement for the neck[20], (mean for left and right) shoulder[21], elbow[22], hip[23], and knee[24] | $a_{i_{t+k/2}} - a_{i_{t-\frac{k}{2}}}, for\ k = 120\ frames$ |
| Amount of movement for the head[25], left and right hand[26,27] and knee[28,29] | $\displaystyle\sum_{l=t-k/2}^{t+k/2} a_{i_l}$ |

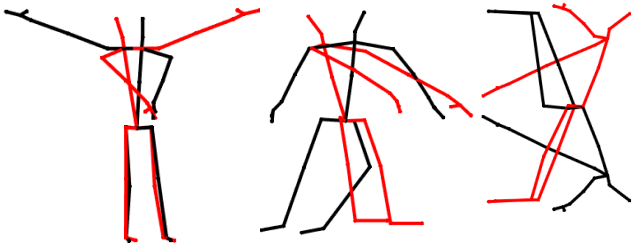

Fig. 5. Skeleton reflection in the x-direction (Left), y-direction (Middle), z-direction (Right): original skeleton in black, reflection in red.

thinking (RT). This resulted in 195 RT and 196 NRT periods for Setting A and 165 RT and 165 NRT periods for Setting

B. Table 2 shows the distribution across task.

### 6.2.2 Data Augmentation

To boost data size, for each period $s_b{}^m$ (where $m$ is the serial index of the period in the corresponding task $b$), we created 3 new periods $_x\bar{s}_b{}^m$, $_y\bar{s}_b{}^m$, and $_z\bar{s}_b{}^m$ whose skeletons are mirror reflections of the skeleton in $s_b{}^m$ in the x, y, and z directions respectively. Mirror reflection augmentation is an approach widely used on image data (e.g. in [16]). In our case, it resulted in four times the original number of periods, i.e. 1564 periods (780 RT and 784 NRT) for Setting A and 1320 periods (660 RT and 660 NRT) for Setting B. Fig. 5 shows examples of our augmented skeletons. Since the behaviour of the skeleton in each period is unaltered in our augmentation, the augmented periods retained the labels of original periods.

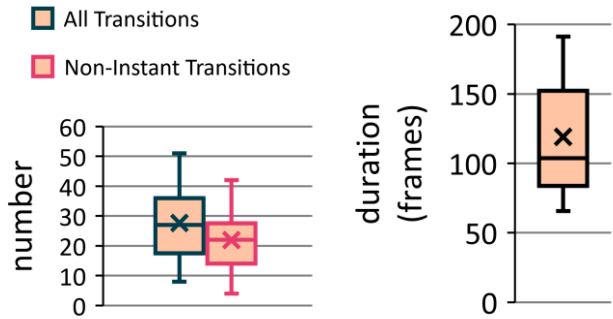

Fig. 6. (a) Left - Mean number (per child) of all orientation transitions and of non-instant orientation transitions in the Forming Angles task; (b) Right - Mean durations of all transitions in this task.

### 6.2.3 Input Forms

In our dataset, each frame $t_{b,m} = 1, 2, …, T_{b,m}$ in each period $s_b^m$ (or $_x \check{s}_b^m$, $_y \check{s}_b^m$, $_z \check{s}_b^m$ for augmented periods) is specified by a set of three-dimensional joint positions $j_{i,t}^{b,m}$ for joint $i$, $i = 1, 2, … 25$. From this set, we extracted two input forms for each period:

Raw Three-Dimensional Positions **POS**: The so-called end-to-end detection (where the learning algorithm directly uses raw sensor data as input) has become increasingly favoured in related areas (e.g. emotion detection in speech [34] and human activity recognition [35]). To explore the possibility of end-to-end detection of reflective thinking based on body movement data, we extracted for each frame $t_{b,m}$, the three-dimensional positions for the major 17 joints. This excluded 8 extremity joints (toes, heels, fingers) that we expected to have more noisy positional data.

Hand-Crafted Features **FEATS**: Further also exploring the traditional method of using informed features for affect detection, we additionally extracted 29 bodily features per frame (see Table 3 for computation formulae) based on the bodily cues discovered from our analysis in Study 2. Prior registration of the data was not needed for this extraction as the features are based on relative metrics. The features include instantaneous energies and range and amount of movement (both computed on a window of up to 60 frames before and after the current frame) to capture pauses, fidgeting, and exploratory movements. The features also include head and trunk orientations to define head and body postures. Hand-to-head distance was additionally included to characterise head self-adaptor behaviours.

### 6.2.4 Dealing with Variation in Period Lengths

The median length of the RT and NRT periods are 114 and 389 frames (interquartile range = 128 and 645.5) respectively for the Setting A data; and 73 and 484 frames (interquartile range = 130 and 1062) respectively for the Setting B data. To make the periods of uniform length, we resampled each period (for both POS and FEATS inputs, augmented and otherwise) to $T=120$ frames. The data had been recorded at 30 frames per second. We expect the resampling approach to introduce less noise than the typical padding method. The choice of $T=120$ is based on findings from an analysis of orientation behaviour in the Forming Angles task in Setting A [37]. The analysis was inspired by findings in [9] that suggest that orientation may offer a glimpse into the attention pattern of a child during a learning activity. In the analysis, we continuously noted the orientation directions for each child all through the task. One of our findings was that transition between orientation targets (e.g. from screen to instructor) was not always instant (Fig. 6a) but rather took about 120 frames (4 seconds) on average (see Fig. 6b) [37]. This finding suggests that 120 frames is a minimum time window on interesting behaviour in the context of the weDraw-1 Movement Dataset.

### 6.3 Results

The results of our experiments on the binary classification of problem-solving periods by reflective thinking are reported in this section under four main themes. Please note that we report accuracy as a proportion (similar to [38] and [39]), rather than as a percentage. We also report F1 score, inverse F1 score, and Matthews Correlation Coefficient [36].

For the neural architectures, input data was scaled to zero mean and unit variance. Further, a gaussian noise (standard deviation = 0.1) was added at the input layers of the networks to limit overfitting.

### 6.3.1 Generalisability of the DD- and V-LSTMNNs

We used leave-one-subject-out cross-validation over 50 epochs with 0.1 learning rate based on a Stochastic Gradient Descent optimizer (momentum set to 0.3) and 30 batches with the Setting A data to evaluate the generalisation ability of the DD- and V-LSTMNNs. These training parameters (and the network parameters) were set based on leave-one-subject-out cross-validated grid search using the Setting A data. The number of trainable parameters for the two architectures were 924 and 105,897 respectively, using FEATS input forms. In our evaluation, we compared the performances of the augmented POS and FEATS input forms.

As shown in upper section of Table 4, there was overall better-than-chance-level performance, with the highest accuracy of 0.75 (average F1 score = 0.75). There was little difference between the performance of the DD- and V-LSTMNNs (accuracy of 0.73 and 0.72 respectively averaged over the two input forms). However, the augmented POS input showed a slight edge over the augmented FEATS input with accuracies of 0.74 and 0.75 compared to 0.71 and 0.69 for the DD-LSTMNN and V-LSTMNN respectively. Augmentation also led to an improvement in performance, albeit a small one, as can be seen in comparing the performances for aug-FEATS and FEATS: accuracy of 0.7 and 0.68 respectively, average for the two NN architectures.

### 6.3.2 Generalisability of Other Standard Algorithms

We compared the performance of the two LSTMNNs with other standard algorithms (CNN, convolutional-LSTMNN, Support Vector Machines, and Random Forest) based on the FEATS input form, also with leave-one-subject-out cross-validation on the Setting A data.

CNNs are more common with data that occur naturally in grid format (e.g. images) [40]. However, the convolution function could be useful in integrating temporal and postural patterns, similar to the LSTM, albeit only locally. The convolutional-LSTMNN combines convolutional and LSTM functions within a single layer. The (hyper)parameters of CNN and convolutional-LSTMNN were based on systematic experimentation with leave-one-subject-out cross-validation. We found two convolution layers followed by two fully connected layers to be optimal for the CNN and three convolutional-LSTM layers followed by one fully connected layer to be optimal for the convolutional-LSTMNN. The same training parameters used for the LSTMNNs were used here, also based on grid search, except for the use of 60 batches in this case. The number of trainable parameters of the CNN and convolutional-LSTMNNs were 533 and 150 respectively.

TABLE 4
OVERALL PERFORMANCE METRIC [LOWER BOUND, UPPER BOUND] VALUES FOR THE CLASSIFICATION MODELS BASED ON LEAVE-ONE-SUBJECT-OUT CROSS VALIDATION
('AUG-' INDICATES INCLUSION OF AUGMENTED PERIODS IN THE DATA; MCC = MATTHEWS CORRELATION COEFFICIENT)

| Algorithm | Input Type | F1 (RT) [0, 1] *chance- level F1 = 0.5* | F1 (NRT) [0, 1] *chance-level F1 = 0.5* | Accuracy [0, 1] *chance-level accuracy = 0.5* | MCC [-1, +1] *chance-level MCC = 0 [35]* |
|---|---|---|---|---|---|
| V-LSTMNN | aug-POS | 0.78 | 0.72 | 0.75 | 0.52 |
| DD-LSTMNN | aug-POS | 0.77 | 0.71 | 0.74 | 0.50 |
| DD-LSTMNN | aug-FEATS | 0.73 | 0.68 | 0.71 | 0.42 |
| V-LSTMNN | aug-FEATS | 0.70 | 0.68 | 0.69 | 0.38 |
| DD-LSTMNN | FEATS | 0.70 | 0.66 | 0.68 | 0.37 |
| Polynomial-kernel SVM | FEATS | 0.70 | 0.66 | 0.68 | 0.36 |
| V-LSTMNN | FEATS | 0.69 | 0.66 | 0.68 | 0.35 |
| CNN | FEATS | 0.66 | 0.60 | 0.63 | 0.27 |
| RF | FEATS | 0.64 | 0.61 | 0.63 | 0.26 |
| Gaussian-kernel SVM | FEATS | 0.67 | 0.19 | 0.53 | 0.12 |
| Convolutional-LSTMNN | FEATS | 0.43 | 0.61 | 0.54 | 0.08 |

TABLE 5
OVERALL PERFORMANCE BASED ON A SEPARATE HELD-OUT TEST SET

| | DD-LSTMNN | | V-LSTMNN | |
|---|---|---|---|---|
| | aug-POS | aug-FEATS | aug-POS | aug-FEATS |
| F1 (RT) | 0.78 | 0.78 | 0.81 | 0.75 |
| F1 (NRT) | 0.75 | 0.70 | 0.77 | 0.70 |
| Accuracy | 0.77 | 0.75 | 0.79 | 0.73 |
| MCC | 0.53 | 0.53 | 0.60 | 0.47 |

TABLE 6
DATA SIZES (TO THE NEAREST HUNDRED) BASED ON PERIOD SUBSEGMENTS

| | $w$ (frames) | | | | | | |
|---|---|---|---|---|---|---|---|
| | 7 | 15 | 30 | 60 | 90 | 120 | 240 |
| Setting A ($\times 10^2$) | 295 | 143 | 76 | 43 | 32 | 27 | 19 |
| Setting B ($\times 10^2$) | 215 | 105 | 57 | 33 | 26 | 22 | 15 |

Further, Support Vector Machine (SVM) [41] and Random Forest (RF) [42] are established as efficacious algorithms for movement-based affect detection (e.g. in [24], [27]). We explored polynomial and gaussian kernels for the SVM in our work. The box constraint of the SVMs were set based on systematic search within the set {10⁻³, 10⁻², 10⁻³, 1, 10, 100, 1000} based on a validation set within a (nested) leave-one-subject-out cross-validation. This was also used to find the optimal degree of the polynomial SVM between 1 and 5. The same approach was applied to find the optimal number of trees (within the set { 50, 100, 200, 500, 1000 }) and the number of features to use to build each tree node (1, all, or a square root of the total number). For the SVM and RF, for each period, the average values (for each feature) over all frames) were used as input. The feature averages were scaled to zero mean and unit variance for both the SVM and RF.

The resulting performances are shown in the lower section of Table 4. The polynomial-kernel SVM had performance similar to the DD- and V-LSTMNNs', with 0.68 accuracy (average F1 score = 0.68). The CNN and RF both performed slightly worse although better than chance level (accuracy = 0.63, average F1 score = 0.63). Both the gaussian-kernel SVM and convolutional-LSTMNN performed poorly with accuracy of 0.53 (0.43 average F1 score) and 0.54 (0.52 average F1 score) respectively.

### 6.3.3 Generalisability of the DD- and V-LSTMNNs Using A Separate Test Set

We further evaluated the generalisability of the DD- and V-LSTMNNs using the Setting A data as the training set and the Setting B data as a held-out test set. This evaluation arrangement enabled us to investigate how well reflective thinking modelling transfers from research lab settings to school settings which is more logistically complex. We used the same training parameters are Section 6.3.1. In this evaluation, we again compared the augmented POS and FEATS input forms.

As can be seen in Table 5, similar to the leave-one-subject-out cross-validation results in Section 6.3.1 based on the Setting A data alone, performance is much better than chance-level detection for the POS and FEATS inputs and for both the DD- (average F1 score = 0.77 and 0.74 respectively) and V-LSTMNNs (average F1 score = 0.79 and 0.73 respectively).

### 6.3.4 Generalisability of the DD- and V-LSTMNNs Based on Subsegments of Event Periods

In a fourth set of experiments, we segmented each RT and NRT period $s_b{}^m$ (or $_x\breve{s}_b{}^m$, $_y\breve{s}_b{}^m$, $_z\breve{s}_b{}^m$ for augmented periods) based on non-overlapping windows of size $w$. End-of-period segments with length less than $w$ were resampled to length $w$. We experimented with $w$ = 7, 15, 30, 60, 90, 120, and 240 frames (the data was recorded at the sampling rate of 30 frames per second). To balance the number of segments from RT and NRT periods, we randomly selected $q$ segments from each NRT period in $S_b$, where $q$ is the mean number of segments resulting from RT periods in $S_b$. Segments with length less than $v = 7$ frames (before resampling) were discarded as too little. The resulting data sizes are shown in Table 6 for each window length. As in above, we used hold-out validation with the Setting A and B data as training and test sets respectively, and we experimented with both augmented POS and FEATS input forms. The results are shown in Fig. 7.

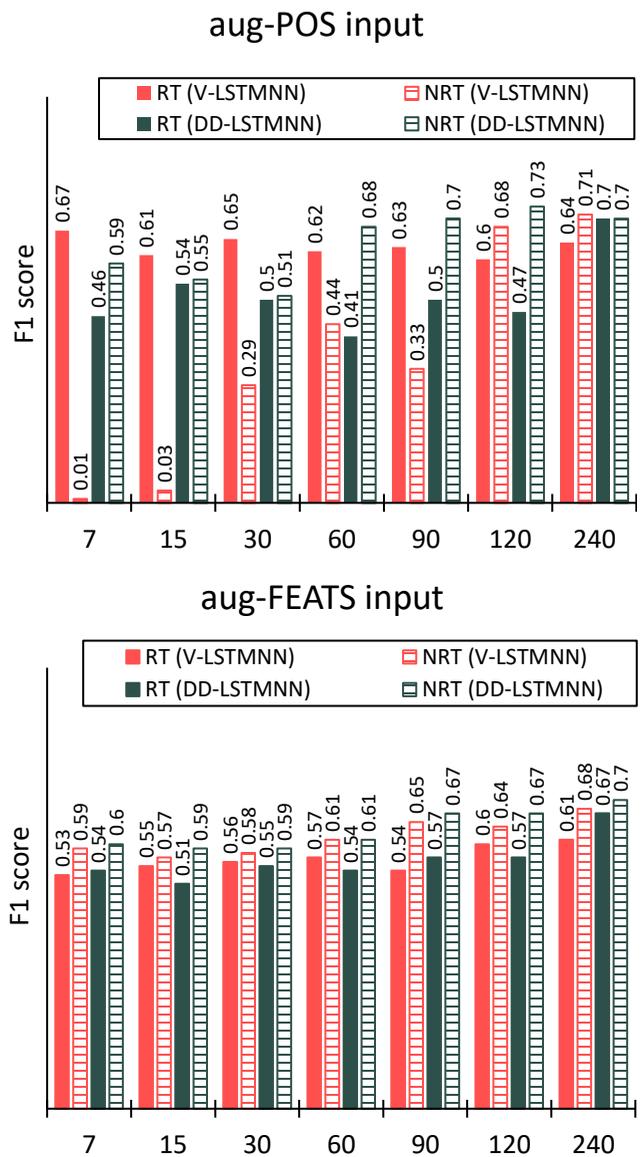

Fig. 7. Overall performance based on 7, 15, 30, 60, 90, 120, and 240-frame subsegments of periods using aug-POS (Top) and aug-FEATS (Bottom) input forms.

For both aug-POS and aug-FEATS input forms, the V- and DD-LSTMNN perform better than chance level detection with $w$=240, giving average F1 scores of: 0.68 and 0.7 for aug-POS (for V- and DD-LSTMNN respectively), and 0.65 and 0.69 for aug-FEATS (for V- and DD-LSTMNN respectively).

For $w$s less than 240, based on the aug-POS input form, the DD-LSTMNN has generally poor performances for the RT class. The V-LSTMNN has even poorer performances (especially for the NRT class, in its own case) for $w$s less than 120 based on the same input form. The better performance of the DD-LSTMNN (albeit not better than chance level) for $w$s less than 120 points to the possibility that delayed integration of the learning across the time and posture dimensions may indeed be a valuable approach to learning signatures of cognitive experiences in motion capture data. For the aug-FEATS input form, both the V- and DD-LSTMNN have comparable performance when $w$ is less than 120. Although these performances are only marginally better than chance level (especially in identifying RT subsegments), they are superior to the performances based on the aug-POS input form for the respective $w$s. This suggests that hand-crafted features are handy (or even necessary) when only (temporally) partial data is available.

# 7 DISCUSSION

Our investigation in this paper was aimed at providing understanding of the feasibility of automatic detection of reflective thinking in children while they solved mathematical problems based on bodily exploration in unconstrained settings. First, we present the weDraw-1 Movement Dataset of body movement of children in these settings, with continuous observer annotation of reflective thinking (as well as low confidence) based on two human raters. The annotated motion capture data from this dataset will be made open (with access via https://doi.org/10.5281/zenodo.2548828) to the research community to facilitate further modelling of reflective thinking and low confidence. Secondly, we provide findings from our analysis of reflective thinking behaviour in this dataset, to inform the extraction of automatic detection features for this state. Comparison of hand-crafted features informed by this analysis to the raw three-dimensional positions from the sensors, in binary classification experiments, suggests that the former may be valuable in moving to continuous detection of reflective thinking. Third, based on both the dataset and our analysis, we showed the possibility of using LSTMNNs for binary classification of problem-solving periods by reflective thinking, with an average F1 scores of 0.73 and 0.79 based on hand-crafted features and raw sensor data respectively. Finally, we explored classification based on period subsegments, toward continuous detection, and showed the feasibility of 0.64 average F1 score for subsegment lengths of 4 seconds and 0.70 average F1 score for lengths of 8 seconds. We discuss our findings in the rest of this section.

## 7.1 Reflective Thinking Behaviours

An interesting finding was that beyond the brief pause that marks reflective thinking, a person in this state is likely to exhibit additional behaviours (e.g. gazing into space, a finger to the chin) that make it recognisable by observers. Our analysis of these behaviours contributes to the oeuvre of bodily action coding taxonomy amassed over a wealth of behaviour analysis literature (e.g. [24]).

Our findings could further inform better understanding of learning processes [15]. We discuss in Section 5.2 the roles of the behaviours found in reflective thinking process. Neuroscientific studies need to be carried out to test our hypotheses or provide empirical theories that explain these findings. In fact, deeper understanding of the functions of these behaviours may lead to more informed design strategies for digital learning technologies. For example, if our hypothesis that gazing into space serves the purpose of removing visual distractors, and so reducing cognition complexity, is accurate, the backlight of the visual feedback screen used could be dimmed during reflective

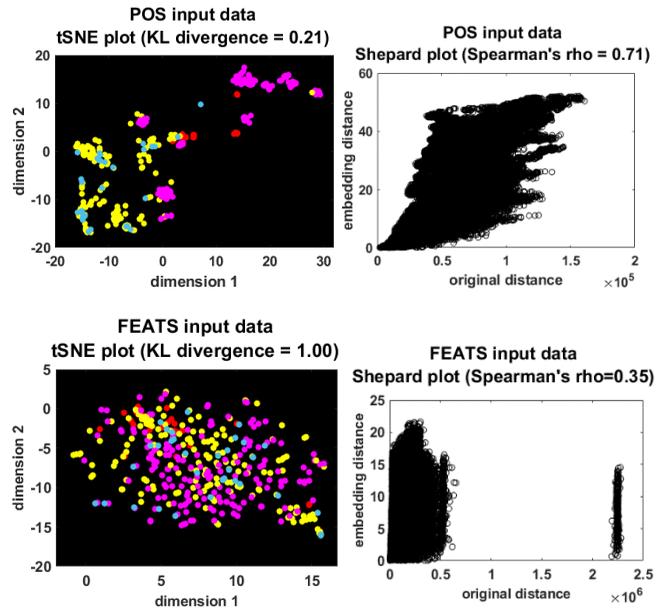

Fig. 8. t-SNE [46] visualisation with corresponding Shepard goodness-of-fit diagrams of Setting A POS (Top) and FEATS (Bottom) input data for: purple - Finding Symmetry & Making Shape Reflections; yellow - Bodily Angles Sums & Rotating in Angles; blue - Bodily Angles Differences & Rotating in Angles; and red - Forming Angles (red).

*KL divergence (Kullback Leibler divergence) ϵ [0, ∞] with optimal value = 0.*

thinking events. Findings in [43] suggest that this or similar approach may support exploration of complex problems which have constituent tasks whose outcomes need to be thoughtfully integrated into the primary solution. Ideally, both internal state (is the child thinking reflectively?) and behaviour (is the child gazing into space?), rather than simply one or the other, should be decoded (as humans do [44]) to deliver appropriate personalisation.

Unsurprisingly as the observers were restricted to visual assessment, the majority of the codes derived for the cues reported by our observers were bodily. However, this finding underscores the value of bodily expressions for recognition of cognitive or emotional states [44] despite the lower attention that it receives in literature, compared to the face. Neuroscientific findings (see [44] for a review) suggest that bodily expressions are in fact given precedence by observers when facial expression is ambiguous. Perhaps this also holds for the case when facial expressions are hard to specify, such as in our findings, with reports like "hard look", "thinking face", "a facial expression I can't describe" by our observers).

## 7.2 Modelling Reflective Thinking using LSTMNNs

The LSTMNN's agreement of 0.60 MCC (average F1 score = 0.79) with the ground truth is similar to the agreement we found between human observers (ICC=0.63, see Section 4.2). This finding is also comparable to findings in previous work on learning-related affect detection [14][16]. Our findings are particularly noteworthy given that the LSTMNN was trained with optical-based motion capture sensor (Kinect) data obtained in much unconstrained settings and so prone to noise due to occlusions, environmental artefact, and being out of camera view.

Further, unlike the human observers in our study who had knowledge about the individual tasks that the children had to complete, the LSTMNN was not explicitly informed that there are several different tasks within the training data. Yet, the model showed very good performance despite the challenge of disentangling behavioural differences due to reflective thinking from those due to task disparity. To support this argument, we apply two-dimensional t-distributed stochastic neighbouring embedding (t-SNE) [46] with euclidean distance to both the POS (Fig. 8 top-left) and FEATS (Fig. 8 bottom-left) input data used for modelling reflective thinking. For the POS data, there indeed emerges clusters related to the different tasks in our dataset suggesting marked dissimilarity between movements performed in the tasks, especially between both Finding Symmetry & Making Shape Reflections (in purple) and the other tasks. This pattern is not so evident with the FEATS data (also note the poor fitness of the embeddings in Fig. 8 bottom-right: Spearman rank correlation $\rho$=0.35, Kullback Leibler divergence=1.00) suggesting, as can be expected, that the features extracted solely to capture reflective thinking have lost information about the tasks. As such, models built on the hand-crafted features as input may struggle to generalize across tasks compared to features learnt by the model directly from the raw sensor data. Our finding of better performance with the raw joint positions data supports this theory, although the crafted features led to performance much better than chance level and were superior when temporally partial data was available. It is expected that training separate models for individual tasks would lead to improvement in detection performance for either input form. The size of our data limited further investigation on this aspect, although it should be noted that our dataset size is comparable to the size of benchmark body movement datasets captured in real scenarios with participants that are not healthy adults (e.g. [47]). We decided not to alternatively include a task identification variable in the input data because of the possibility of a task effect for reflective thinking in the dataset. The resampling done to force the input data to be of uniform sizes may also have contributed to the generalisation difficulty for the LSTMNN. Nevertheless, the results in Study 3 show that using the complete period (even when resampled) gives better results than the use of windows within these periods.

The fact that the raw joint positions led to better performance moreover highlights the capability of the LSTM layers, despite the low depth of our LSTMNNs and the limited size of our dataset, to learn useful features of reflective thinking itself from raw motion capture data. This discovery makes it more pertinent to create specialised algorithms tailored to motion capture data configurations and so realise the ambition for end-to-end detection [48] for higher-order action properties like cognitive or emotional states. Although preliminary, further findings with the DD-LSTMNN suggest that multiscale architectures (such as in [15]), where learning gradually flows (from the input to output layers) from the component to the whole, is a promising approach for this direction. Intricate network architecture designs such as in [17] can enable higher detection performances, and perhaps even lead to deeper understanding of how components (e.g. anatomical joints) contribute to the expression of the given cognitive or emotional state.

Comparing LSTMNNs with other algorithms, we found the SVM (with polynomial kernel) to closely match the performance of the LSTMNN. This finding is not surprising given previous findings in [24], [27] (amongst others) with the SVM. However, we hypothesize that an increased data size would enable more intricate representations of the temporal relations in movement behaviour in the LSTMNN and so give it an edge over the SVM.

## 8 CONCLUSION

Automatic detection of reflective thinking episodes in a child solving a mathematical problem is a pertinent problem for tailored feedback/support in technology for mathematics learning. The findings of this paper demonstrate the feasibility of binary classification that matches human assessment in this context, with average F1 score of 0.79. Further, it is shown that 4 seconds may be adequate for automatic recognition of a reflective thinking episode. Our new weDraw-1 Movement Dataset captured in real, unconstrained settings is the basis for these conclusions. Human annotations show the presence of episodes of reflective thinking in the dataset, with good level agreement between two raters. In their assessment, the human raters were found to use cues that the limited existing literature suggest as associated with the different phases of reflective thinking. These outcomes highlight the opportunity for learning technology to recognise moments of reflective thinking in unconstrained settings. Our dataset and analysis also pave the way for further work on continuous detection of reflective thinking as well as automatic detection of facets of the states (e.g. discriminating between the phase of idea generation and evaluation), additionally highlighting relevant questions for related disciplines.

### ACKNOWLEDGMENT

This project has received funding from the European Union's Horizon 2020 research and innovation programme under grant agreement No 732391.

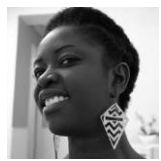

**Temitayo A. Olugbade** is an applied ML/AI researcher and currently a Research Associate at University College London. Her interests are in affective intelligence (i.e. building psychological models of humans for AI systems) to enrich human-AI interaction/collaboration and AI in general. She has a PhD in Affective Computing, with background in AI (MSc), and Computer Engineering (BSc).

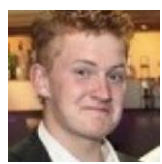

**Joseph Newbold** is a PhD student at UCL and part-time researcher on the weDraw project. His research interests are based on how sound can be used to improve people's interaction with technology and alter their perception of the world and themselves.

**Rose Johnson** has a PhD in Human Computer Interaction from UCL and a background in Physics. Her areas of interest are educational technology, whole body interaction and multimodal feedback. She is helping on the weDraw project as a visiting researcher while looking after her young family.

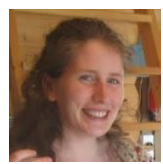

**Erica Volta** received her M.Sc. degree in Cognitive Science and Decision-making Process. She is currently a Ph.D. fellow in Informatics and System Engineering at the University of Genoa, Italy. Her research interests include adaptive and affective human-machine interaction, particularly applied to multisensory and multimodal integration design in learning and rehabilitation contexts, and technology integration with performing arts (mainly music). For her PhD, she is currently involved in two H2020 projects (TELMI, weDRAW).

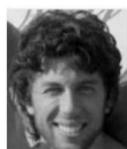

**Paolo Alborno** received his M.Sc. degree in Computer Engineering. He is currently a Ph.D. fellow in informatics and system engineering at the University of Genoa, Italy. His research interests include intelligent and affective human-machine interaction, autonomous systems, modeling and real-time analysis and synthesis of expressive content, and multimodal interactive systems. is currently involved in various in H2020 projects (DANCE, WHOLODANCE, TELMI, WeDRAW).

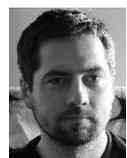

**Radoslaw Niewiadomski** received the PhD degree in Computer Science from the University of Perugia, Italy. His research interests are in the area of affective computing and include: recognition of emotions, nonverbal behavior synthesis, creation of embodied conversational agents and interactive multimodal systems. He has been involved in several FP6, FP7 and H2020 EU research projects among which are FP6 CALLAS, FP7 ILHAIRE and H2020 DANCE. He has published over 50 peer-reviewed conference and journal papers. He was the co-editor of the Special Issue dedicated to laughter computing of IEEE Transactions on Affective Computing.

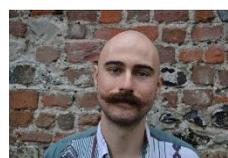

**Max Dillon** is a data scientist on the central AI team at PwC. He holds a MSc in Machine Learning and Computational Statistics from University College London, where he conducted his dissertation research with the Human-Robot Interaction group, building an automatic classification system for pain level in people with chronic lower back pain. Previously, he studied Economics at the University of Oxford.

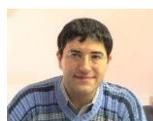

**Gualtiero Volpe** received the M.Sc. degree in computer engineering in 1999 and the Ph.D. in electronic and computer engineering in 2003 from the University of Genoa, Italy. Since 2014, he is an Associate Professor at DIBRIS, University of Genoa. His research interests include intelligent and affective human-machine interaction, social signal processing, sound and music computing, modeling and real-time analysis and synthesis of expressive content, and multimodal interactive systems.

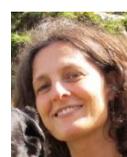

**Nadia Bianchi-Berthouze** is professor in Affective Computing and Interaction. Her main area of expertise is the study of body posture/movement as a modality for recognising, modulating and measuring human affective states in human-computer interaction. She has published more than 170 papers in affective computing, human-computer interaction, and pattern recognition. She was awarded the 2003 Technical Prize from the Japanese Society of Kansei Engineering and she has been invited to give a TEDxStMartin talk (2012). She is/was: PI on the EPSRC-funded Emo&Pain project to design affective technology to support rehabilitation in chronic pain; co-I on the EPSRC-funded Digital Sensoria project investigating the use of biosensors to measure subjective responses to tactile experiences; co-I on the ILHAIRE project investigating laughter in human-avatar interaction; EU-FP7 Marie Curie IRSES UBI-HEALTH: Exchange of Excellence in Ubiquitous Computing Technologies to Address Healthcare Challenges, H2020 HUMAN Manufacturing, and HOLD funded by the Wellcome Trust.